\documentstyle[epsf,psfig,prl,multicol,aps]{revtex}
\begin{document}
\title{Extreme Electron-Phonon Coupling
  in Boron-based Layered Superconductors} 
\author{J. M. An,$^1$ S. Y. Savrasov,$^2$ H. Rosner,$^3$ and
W. E. Pickett$^3$}
\address{
$^1$Lawrence Berkeley National Laboratory, Berkeley CA 94720  \\
$^2$Department of Physics, New Jersey Institue of Technology, Newark, 
    NJ 07102 \\
$^3$Department of Physics, University of California, Davis CA 95616 \\
}

\date{\today}
\maketitle
\begin{abstract}
The phonon-mode decomposition of the electron-phonon coupling in the
MgB$_2$-like system Li$_{1-x}$BC is explored using first principles
calculations.
It is found that the high temperature superconductivity of such systems
results from extremely strong coupling to only $\sim$2\% of the phonon modes.
Novel characteristics of $E_{2g}$ branches include 
(1) ``mode $\lambda$'' values of 25 and 
greater compared to a mean of $\sim 0.4$ for other modes,
(2) a precipitous Kohn anomaly,
and (3) $E_{2g}$ phonon linewidths within a factor of $\sim$2 of the
frequency itself, indicating impending breakdown of linear 
electron-phonon theory.   This behavior in borne out by recent inelastic 
x-ray scattering studies of MgB$_2$ by Shukla {\it et al.}

\end{abstract}

\begin{multicols}{2}
Superconductivity near T$_c 
\approx$ 40 K in
MgB$_2$~\cite{akimitsu} has necessitated a re-evaluation of our 
understanding of phonon-coupled superconductivity, and illustrated
vividly that there has been a substantial void in our 
conceptual understanding of
this ``conventional" mechanism of coupling, long thought to require $d$
electrons, high density of states, and high symmetry.  The general theory 
of electron-phonon (EP) coupled superconductivity has
long been available,\cite{SSW,PBA} and some essential 
understanding of the mechanism in MgB$_2$
is now in place.
\cite{jan,kortus,kong,bohnen,yildirim,liu,berkeley}
The particular structure and chemistry of MgB$_2$ creates 
holes in the B $2p\sigma$ bands,
and these holes are coupled very strongly to the B-B bond stretching
vibrational modes.  The predictions of band theory have been verified
by angle-resolved photoemission\cite{arpes} and de Haas -- van Alphen 
measurements.\cite{dhva,dhva1,dhva2}
One new facet is an unusually strong anisotropy in
the EP coupling, such that the Fermi surface (FS) averaged 
EP coupling strength $\lambda$ and Eliashberg function $\alpha^2 F(\omega)$
do not give a precise account of T$_c$ or of thermal and spectroscopic
data.  It seems however that much can be accounted for by generalizing to
a two-band model\cite{liu,berkeley} of strongly 
coupled $\sigma$ holes 
and weakly coupled $\pi$ electrons,
with $\lambda = \lambda_{\sigma} + \lambda_{\pi} \approx 0.6-0.8$. 

In spite of success in accounting for many observations, this linear EP
theory (with some {\it ad hoc} correction for anharmonicity) seems at
best incomplete.  The enormous linewidth $\gamma$ of the $E_{2g}$ mode,
\cite{hlinka,martinho,quilty,rafailov}
ascribed to strong EP coupling plus anharmonicity, has not been accounted
for quantitatively, and the predicted change in frequency and $\gamma$
at T$_c$ \cite{liu} is not observed.  
Non-adiabatic processes arising from the small filling of the $\sigma$ 
bands, and low Fermi energy E$_F$, have been proposed.
\cite{liu,cappelluti,son}
Boeri {\it et al.} provided needed clarification
by demonstrating that $E_{2g}$ anharmonicity arises 
in MgB$_2$ from non-adiabatic
effects,\cite{boeri} due to the huge deformation
potential.\cite{jan}

A second compound in the ``MgB$_2$ class'' that has been proposed recently
is useful in clarifying these issues.
LiBC is isostructural\cite{worle} and isovalent with MgB$_2$, 
having graphene-like B-C layers that are even more strongly
bonded than are the B layers of MgB$_2$.  The stoichiometric compound
is semiconducting (due to the B-C difference), but Li deficiency that
makes it a hole-doped metal has been reported.\cite{worle}  
Such hole doping introduces carriers into the B-C $\sigma$ bands,
as occurs in MgB$_2$, and near rigid-band behavior has
been demonstrated, making it a promising system for probing the carrier
density dependence of EP coupling in MgB$_2$-class superconductors 

In this paper 
we focus on the 
decomposition of $\lambda$ into contributions (``mode $\lambda$'s'') from 
each phonon $Q\nu$ rather than its FS decomposition, and find
startlingly high values $\lambda_{Q\nu}\approx$ 25 for $E_{2g}$ modes.
First principles calculations of $\alpha^2 F$ for Li$_{0.75}$BC gives
$\lambda$=0.74, making Li$_{1-x}$BC at this doping level quantitatively
similar to MgB$_2$.
The insulating end member of the  Li$_{1-x}$BC system makes it
pedagogically an ideal material to illustrate
the novelty that MgB$_2$ has introduced into the physics of electron-phonon
coupled superconductivity.  Li$_{1-x}$BC is ``MgB$_2$ with stronger coupling
and variable carrier concentration,'' and because its 
$\sigma$ FSs are (as in MgB$_2$) very close to cylindrical,
the underlying physical processes can be modelled simply yet
realistically.  The essential revelations are: 
(1) the high temperature superconductivity
arises from exceedingly strong coupling of only a small fraction
($\sim$2\%, depending on doping) of the phonon modes, 
(2) a colossal sharp Kohn anomaly
occurs -- extreme renormalization by EP coupling upon doping from
hard bond-stretching modes, and
(3) the phonon
linewidth due to decay into $\sigma$ band electron-hole pairs 
becomes comparable to the linewidth itself, making the bond-stretching
modes poorly defined (at best), yet not unstable (due to their intrinsic
hard nature).  Just this strength of coupling and relative linewidths have
been reported for MgB$_2$ by Shukla {\it et al.}, obtained from inelastic
x-ray scattering spectrosciopy.\cite{shukla}
Moreover, (4) these effects occur continuously with doping 
and will allow a unique window into the non-adiabatic regime
where the Fermi energy is comparable to the phonon frequency.  These
features apply directly to Mg$_{1-x}$Al$_x$B$_2$ as well.

\begin{figure}[tbp]
\epsfxsize=7.0cm\centerline{\epsffile{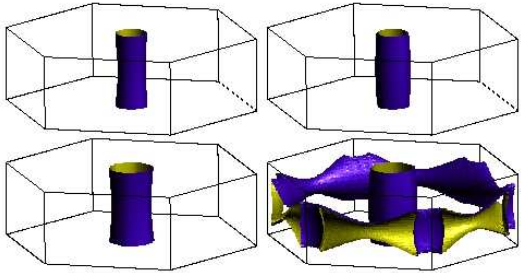}}
\caption{Fermi surfaces of Li$_{0.75}$BC, with the $\Gamma$ point at the
center of the hexagonal zone.  There are four cylinders, 
analogous to the two cylinders in MgB$_2$ but downfolded due to the
doubled unit cell.  The non-cylindrical surface arises from the 
weakly coupled $\pi$ bands.}
\label{fig1}
\end{figure}

A description of the band structure of Li$_{1-x}$BC was presented 
earlier.\cite{LiBCucd}  The FSs for $x$=0.25
pictured in Fig. \ref{fig1}, consist of four B $2p$ -- C $2p$ $\sigma$
band cylinders with very small $k_z$ dispersion, and a $\pi$ band FS.
The $\sigma$ cylinders are similar to those in MgB$_2$, with four
(instead of the two in MgB$_2$) arising from the doubling of the unit
cell along $c$ due to alternate stacking of B-C layers.
The phonon energies and 
EP matrix elements have been obtained from linear
response theory,\cite{linear} as implemented in Savrasov's 
full-potential linear
muffin-tin orbital code.\cite{Sav,LMTO}  Because of the emphasis here on 
specific Fermi surface effects, a dense grid of Q points was 
chosen (a 16,16,4 grid giving 90 Q points in the irreducible wedge).
For k-space integration a finer 32,32,8 grid was used, together
with an adaptive tetrahedron integration scheme.\cite{Sav}  The code was used
previously for MgB$_2$ by Andersen's group,\cite{kong} where the need
for careful zone sampling was also emphasized.  

\begin{figure}[tbp]
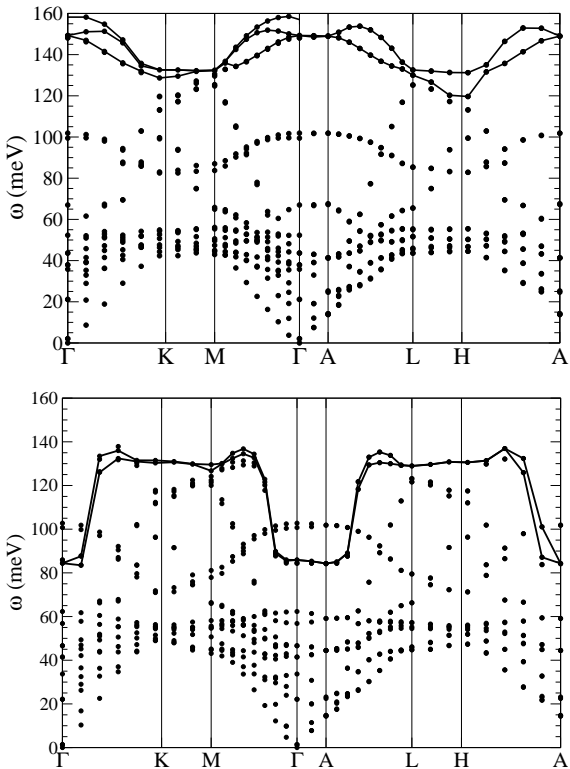

\begin{center}
\psfig{figure=figure2a.eps,width=7.5cm,angle=-0}
\vskip  3mm
\psfig{figure=figure2b.eps,width=7.5cm,angle=-0}
\end{center}
\vspace{3mm}
\caption{
Calculated phonon dispersion curves for (top) semiconducting LiBC 
and (bottom) strongly hole-doped and metallic
Li$_{0.75}$BC, with $E_{2g}$-derived modes connected by heavy lines.  
The primary difference is the extremely strong
renormalization downward ($\omega_Q^2$ decreases by $\sim$60\%) for
$Q < 2k_F$; the extreme van Hove singularities along $\Gamma - K, 
M - \Gamma, A - L, H - A$ are apparent.
}
\label{fig2}
\end{figure}

For the semiconducting phase, the E$_{2g}$ phonons (Fig.\ref{fig2}) are 
very hard, lying at 150 meV at the $\Gamma$ point and not dispersing 
greatly.  This high bond-stretching frequency, twice as high as in MgB$_2$,
is due to the much stronger B-C bond,
9\% shorter and having a 40\%
larger deformation potential for the $\sigma$ bands.\cite{LiBCucd}  
The spectrum is indicative
of a stable, strongly bonded material.  For $x$ = 
0.25, where it has become a metal (Fig.\ref{fig2}),
there is rather little difference in the spectra, 
{\it except} for $Q_{\parallel}$ $ < 2 k_F \approx$ $\pi/3a$ 
($Q_{\parallel}$ is the
component of the phonon wavevector $\vec Q$ lying in the plane).
Within this region, however, the four $E_{2g}$-related modes display
a precipitous Kohn anomaly at $2k_F$, and dip {\it nearly discontinuously by
40\%}.  (The decrease in $\omega^2$, which directly reflects
the renormalization, is nearly 65\%).  Closely related renormalization
of phonon modes in the Mg$_{1-x}$Al$_x$B$_2$ system has been noted
by Renker {\it et al.}\cite{renker}

\begin{figure}[bt]
\begin{center}
\begin{minipage}{7.5cm}
\psfig{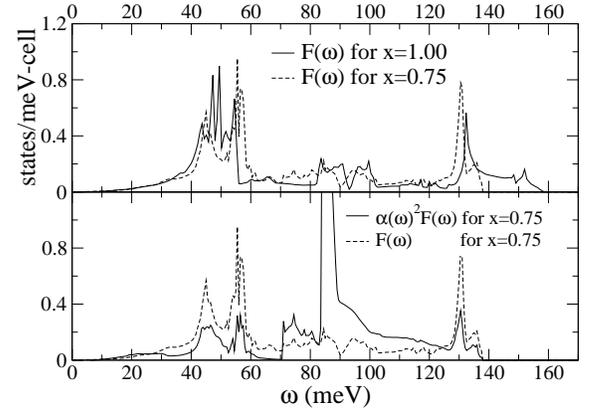}
\end{minipage}
\end{center}
\vspace{3mm}
\caption{
Phonon DOS for LiBC and Li$_{0.75}$BC (top panel)
where the primary difference is the disappearance of modes in the
range 135 -- 155 meV upon doping (from Fig. 1, they mostly lie in the
80-90 meV region).  The shape of $F(\omega)$ and $\alpha^2 F(\omega)$
for Li$_{0.75}$BC (bottom panel) revealing the extremely strong 
coupling to phonons in the 85 -- 100 meV range.
}
\label{fig3}
\end{figure}

In Fig. \ref{fig3} the phonon density of states (DOS) for $x$ = 0 and $x$
= 0.25 are contrasted, and also the calculated EP spectral function
$\alpha^2 F(\omega)$ for $x$ = 0.25 is compared with the DOS.  For the
DOS, doping moves the modes in the frequency range 135 - 155 meV to lower
frequency; only from the dispersion curves of Fig. \ref{fig2} is it
clear where they end up (in the 85 meV region) because these modes are
near the $\Gamma$ point where phase space is small.  The
curve for $\alpha^2 F(\omega)$ demonstrates what results: a
tiny fraction of modes
(quantified below) in the 85 meV region are extremely strongly
coupled to the $\sigma$ band holes, and leads to
an inordinately strong contribution to $\lambda$, whose total value 
calculated from $\alpha^2 F$ is
0.74.  
Although about half of the coupling arises from all other modes,
it is the extremely strong coupling of small-Q E$_{2g}$ modes
to $\sigma$ holes that drives T$_c$, just as
in MgB$_2$.  The calculated value of T$_c$ using the Allen-Dynes
equation\cite{dynes} is
34 K assuming $\mu^*$ = 0.09 as has been used
previously\cite{berkeley}
to obtain numerical agreement with the observed T$_c$ for MgB$_2$.
Solution of the anisotropic Eliashberg equations\cite{berkeley}
would give T$_c$
a few degrees higher.

The circular $\sigma$ Fermi surfaces of MgB$_2$ and related materials
allow an analytic treatment
of EP coupling.  
The phonon energies ($\hbar$=1) 
are given by the poles of the lattice Green's function
\begin{eqnarray}
\omega^2_{Q \nu} = \Omega^2_{Q \nu} + 2 \Omega_{Q \nu} 
              \Pi^{\sigma}(Q,\omega_{Q\nu}).
\label{selfe}
\end{eqnarray}
where the reference frequency $ \Omega_{Q \nu}$ 
includes all self-energy effects
except those arising from electron scattering within the $\sigma$ bands;
practically speaking, they are the energies shown in Fig. \ref{fig2} for
undoped LiBC.  Taking for simplicity a single cylindrical Fermi
surface with N($\varepsilon$)=$m^*/2\pi$ per spin per unit area, 
and considering the
$E_{2g}$ derived modes only,
the phonon self-energy
due to the $\sigma$ carriers is ($\eta_Q \equiv Q_{\parallel}/2k_F$)
\begin{eqnarray}
\Pi^{\sigma}(Q,\omega)=-2 \sum_k |M_{k,Q}|^2 
            \frac{f_k -f_{k+Q}}{\varepsilon_{k+Q}
   -\varepsilon_k -\omega -i\delta} \nonumber  \\
\Re e~ \Pi^{\sigma}(Q,\omega_Q)   
          \approx -2|M_{2g}|^2 N(\varepsilon_F) \hat{\chi}^{2D}_L(Q),\\
 \hat{\chi}^{2D}_L(Q)=\theta(1-\eta_Q) + 
   \bigl[1-\sqrt{1-\eta_Q^{-2}}\bigr]\theta(\eta_Q - 1).
\label{eqn234}
\end{eqnarray}
Here $f_k \equiv f(\varepsilon_k)$ is the Fermi occupation factor, and 
the usual adiabatic approximation has been made.  
A mean square matrix element has been extracted from the sum, leaving the 
unitless Lindhard function $\hat{\chi}^{2D}_L$.\cite{lindhard}
The negative of $\hat{\chi}^{2D}_L(Q)$ has much the form that is
evident in the softened E$_{2g}$ branches with $Q < 2k_F$ in the
bottom panel of Fig. 2.  This behavior is illustrated by plotting
$\omega_Q$ given by Eqs. \ref{selfe}-\ref{eqn234} in the bottom panel
of Fig. 4.

The ``mode $\lambda$'' $\lambda_{Q\nu}$\cite{allen} for the $E_{2g}$
modes, defined such that it is an intensive quantity whose
{\it average} over the zone
and over the N$_{\nu}$ branches gives $\lambda_{2g}$
($\lambda = \lambda_{2g} + \lambda_{other})$, is
\begin{eqnarray}
\lambda_{\vec Q} &=& \frac{2 N_{\nu}}{\pi N(\varepsilon_F)} 
  \frac{\gamma_Q}{\omega_Q^2},
\gamma_Q = \pi \Omega_Q |M_{2g}|^2 \xi(Q), \nonumber \\
\xi(Q)&=&\sum_k \delta(\varepsilon_k) \delta(\varepsilon_{k+Q})
  \equiv N(0)^2 \frac{A_{BZ}}{A_{FS}} \frac{\hat \xi(Q/2k_F)}{\pi}
\label{lambda}
\end{eqnarray}
where $\hat \xi(\eta) = \eta^{-1}(1-\eta^2)^{-1/2} \theta(\eta) \theta(1-\eta)$.
Here $A_{BZ}$ is the basal area of the
BZ and $A_{FS} = \pi (2k_F)^2$ is a doping-dependent area of renormlalized
$E_{2g}$ phonons.
$Q\rightarrow 0$ requires somewhat more care\cite{COR} but involve 
negligible phase space for the present purposes.
The important feature, aside from the magnitude, is that these 
mode $\lambda$'s {\it scale inversely with
$k_F^2$} as illustrated in Fig. 4, diverging for vanishing 
$\sigma$ hole doping and indicating the
breakdown of conventional EP theory (see below).

\begin{figure}[tbp]
\psfig{figure=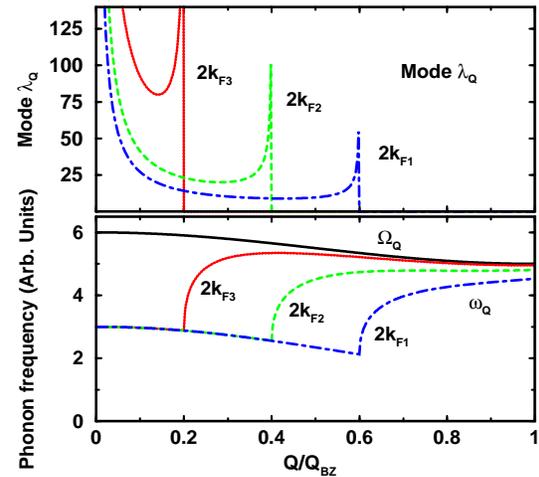,width=7.0cm,angle=-00}
\caption{Characteristics of the model 2D electrons (see text) strongly
coupled to $E_{2g}$ modes.  Bottom panel: $E_{2g}$ branch versus Q;
reference branch $\Omega_Q$ and renormalized branch $\omega_Q$ as in
Eqs. (1)-(3), for three different values of $2k_F$.  Top panel: 
Corresponding values of the mode $\lambda_Q$ given by Eq. \ref{lambda}, 
for the same three values 
of $2k_F$.
}
\label{fig4}
\end{figure}

For $Q < 2k_F$, $\omega_Q$ and 
$\Omega_Q$ are nearly constant, so
Eqs. (1) and (2) can be solved for $|M_{2g}|^2$
and substituted
into Eq. (4) to provide a simple expression for the mode $\lambda_{Q}$ 
($Q < 2k_F$) in terms of $(\Omega^2_{2g} - \omega^2_{2g})/\omega^2_{2g}$.
Similar expression have been used for Nb-Mo alloys\cite{weppba}
and for Pb alloys\cite{pbarcd} to relate $\lambda$ to phonon softening.
There is a simpler way to obtain
the average mode $\lambda$ for E$_{2g}$ phonons.  The total contribution is
that obtained from the 85-100 meV region (see Fig. \ref{fig3})
of $\alpha^2 F$, $\lambda_{2g}$
=0.4.  This arises from an average over all phonons, but only those in
3/40 of the BZ ($Q < 2k_F$) and the upper four of 18 of the branches contribute.
Hence the mean is $<\lambda_Q^{2g}>$ = 0.4$\times (40/3)\times (18/4)$
$\approx$ 25.  The full $Q$ dependence of
the mode $\lambda_{Q}$ for the $E_{2g}$ branches is
given by Eq. (4) and shown in the top panel of Fig. \ref{fig4}.
The mean mode $\lambda$ from all other
phonons (equal to their contribution to $\lambda$ since it includes 98\%
of all modes) also is $\lambda_{other} \approx$ 0.4, accounting for the
total $\lambda = 0.74$ obtained from $\alpha^2 F$. 

The phonon relative halfwidth\cite{allen} becomes
$\gamma_Q/ \omega_Q 
\sim 0.5 \lambda_Q/<\lambda_{2g}> \sim 0.5$,
an alarming result (this ratio is typically 10$^{-2}-10^{-3}$)
that reflects the fact that this phonon branch is so
ill-defined from extremely strong EP coupling that Migdal theory
(and Eliashberg theory) is no longer justified.
Even so, this rough estimate is consistent with the large observed  
halfwidth\cite{hlinka,martinho,quilty,rafailov} 
of 125-175 cm$^{-1}$ for the 600 
cm$^{-1}$ bond-stretching mode, indicating that the large linewidth and
its strong increase with temperature is due to EP coupling
rather than to anharmonicity, and the same conclusion has been reached
by Shukla {\it et al.}\cite{shukla}

Now we summarize and discuss some implications of these results.  Use of
Li$_{1-x}$BC has allowed us to identify and quantify 
the drastic phonon softening arising from
the ultrastrong EP coupling to $E_{2g}$ modes; 
comparable phenomena occur also in 
MgB$_2$.  Mode $\lambda_{2g}$'s $\approx$ 25 and linewidths comparable to
the frequency point to
inadequacies of linear EP theory for these systems.  
Shukla {\it et al.} have reported\cite{shukla} measurements of 
specific frequencies and linewidths of MgB$_2$
from inelastic x-ray scattering. 
For five small-$Q_{\parallel}$ E$_{2g}$ phonons with frequencies in
the 500-550 cm$^{-1}$ range, they obtained relative linewidths
$\gamma_Q/\omega_Q \approx$ 1/3, very consistent with our results
above.  For these modes, the average $\lambda_Q$ = 26 (when normalized
as we have done) are again exactly in line with the expectations
outlined above.

This extremely strong and abruptly Q dependent EP coupling provides new
insight into the relationship of limits on T$_c$ and lattice
instability.\cite{instab}  Earlier study had produced arguments both
that incipient lattice instabilites are helpful\cite{weppba,pbarcd} 
for T$_c$, and 
that they are unhelpful because the low frequency modes are less
useful for high T$_c$ than higher frequencies 
(of the order of $2\pi$T$_c$).\cite{rainer}
The lack of explicit dependence of $\lambda_{2g}$
on $k_F$ 
indicates that increase in T$_c$ with doping will arise only
from an increase of $m^*$ (the heavy hole mass increases by 25\% from
$x=0$ to $x=0.5$) or increase of $|M_{2g}|^2$, either of which will 
further soften the $E_{2g}$
modes and move the system closer to instability.
What changes rapidly with doping is the coupling {\it strength} 
of the $E_{2g}$ modes
with $Q < 2k_F$ --
the same total EP coupling strength is concentrated 
into increasingly fewer bond-stretching modes at lower doping levels.
Thus in this system (where the two dimensionality of the electron
dispersion is central) change of the doping level can lead toward (or
away from) a lattice instability without any direct affect on T$_c$
itself.
These new features involving extremely strong coupling apply
to Mg$_{1-x}$Al$_x$B$_2$, where unexplained structural anomalies
occur, as well as to Li$_{1-x}$BC.

W.E.P. acknowledges important discussions with I. I. Mazin and O. K. Andersen
related to mode $\lambda$'s in MgB$_2$.
This work was supported by National Science Foundation Grant
DMR-0114818, and by the Deutscher Akademischer Austauschdienst.

\vskip -7mm

\end{multicols}
\end{document}